\documentclass[preprint,12pt,authoryear]{elsarticle}

\usepackage{amssymb}
\usepackage{amsmath}
\usepackage{graphicx}
\usepackage{float}
\usepackage{caption}
\usepackage{subfig}
\usepackage{url}
\usepackage{hyperref}
\hypersetup{colorlinks=true, urlcolor=cyan}
\begin{document}

\begin{frontmatter}



\title{Valeriepieris Circles Reveal City and Regional Boundaries in England and Wales} %% Article title


\author[inst1]{Rudy Arthur\corref{cor1}} 
\ead{R.Arthur@exeter.ac.uk}
\cortext[cor1]{Corresponding author}
\author[inst1]{Federico Botta}

\affiliation[inst1]{organization={University of Exeter, Department of Computer Science},%Department and Organization
            addressline={North Park Road}, 
            city={Exeter},
            postcode={EX4 4RN}, 
            country={UK}
}

%% Abstract
\begin{abstract}
We propose a new method of determining regional and city boundaries based on the Valeriepieris circle, the smallest circle containing a given fraction of the data. 
By varying the fraction in the circle we can map complex spatial data to a simple model of concentric rings which we then fit to determine natural density cutoffs. We apply this method to population, occupation, economic and transport data from England and Wales, finding that the regions determined by this method affirm well known social facts such as the disproportionate wealth of London or the relative isolation of the North East and South West of England. We then show how different data sets give us different views of the same cities, providing insight into their development and dynamics.
\end{abstract}



%% Keywords
%\begin{keyword}



%\end{keyword}

\end{frontmatter}



\section{Introduction}\label{sec:intro}

The spatial distribution of people, infrastructure, resources, languages or any social variable reflects the complex history of a landscape since its settlement. Despite this complexity there is enough regularity in the relationships between variables to support the idea of a \textit{science of cities}~\citep{batty2012building} with a number of basic statistical laws \citep{zipf2016human, gibrat1931inegalits, bettencourt2007growth,bettencourt2010unified}. A science of cities clearly depends on how we define what is and is not in a given city. In fact some of these laws, especially allometric scaling laws \citep{bettencourt2007growth}, are still subject to debate, with some arguing that they depend on particular definitions  and characterisations of a city, see especially \cite{arcaute2015constructing} and subsequent work \citep{cottineau2015paradoxical, leitao2016scaling, rybski2019urban}.

One can rely on political or administrative boundaries to define a city, however election districts can be gerrymandered to favour some political party and administrative boundaries may be 
historic, failing to reflect the reality on the ground \citep{batty2011defining,taubenbock2019new}. Due to this a number of data driven approaches have been proposed, offering a more objective definition of a city or region. The City Clustering Algorithm (CCA) \citep{rozenfeld2008laws,rozenfeld2011area} joins populated grid squares, identifying connected clusters with cities. \cite{arcaute2016cities} study percolation on street networks to identify cities and regions. Other approaches exist based on fractal geometry \citep{tannier2011fractal}, population growth modelling ~\citep{makse1995modelling} or clustering algorithms \citep{pavlis2018modified,arribas2021building} and all seek to determine city boundaries directly from data such as census counts~\citep{arcaute2015constructing, corral2020truncated}, satellite imagery~\citep{bennett2017advances,xu2021mapping}, or land-use maps \citep{cieslak2020use}.

As cities grow, merge and become increasingly interlinked, considering them in isolation may not be appropriate \citep{batty2023boundary}. Regions, comprising multiple cities, may be better units of analysis but determining regional boundaries is just as challenging. We take Britain as our primary example. As an extremely densely populated island with a long history of urbanisation, Britain is a case study in problems of boundary definition. Many British cities are the result of mergers of historically separate towns, a process which is still occuring e.g. the city of Brighton-Hove was officially designated in 2001. Other cities are becoming less distinct members of larger conurbations e.g. Greater Manchester Built-up Area, Leeds-Bradford. At the regional level Wales is a separate country of 3.2 million people, with a distinct political assembly and language, but with no restriction on travel between Wales and England. There is also the North/South divide in England, a cultural, political, economic and even linguistic \citep{grieve2019mapping} division that is generally recognised but with many different definitions of `north' and `south' e.g. including or excluding the Midlands \citep{bambra2023northern}. Scotland is increasingly politically separate from England and Wales and conducts its own census. For this reason we restrict our analysis to England and Wales (abbreviated EW) and also exclude Northern Ireland.

A data driven approach to areal demarcation should reflect these known divisions without imposing them. The closest work to ours is  \cite{arcaute2016cities} who study exactly the problem of determining the cities and regions of England and Wales from data. By studying percolation clusters of the road network at various thresholds, a natural hierarchy emerges, first of cities then of regions. Determining the right threshold that identifies, say, cities, is not straightforward. \cite{molinero2021geometry} use a heuristic based on maximising the population-weighted, average fractal dimension of the clusters while \cite{masucci2015problem} use an approach based on determining the `condensation threshold'. The use of the road network is well justified by \cite{arcaute2016cities} but different data might give rise to different boundaries e.g. rail rather than road infrastructure, population density or economic activity could all lead to different definitions of a city or region \citep{paul2017limits}. If we want to use population or economic data in a network percolation setting, it must first be triangulated  \citep{molinero2021geometry}. A simpler approach to boundary finding is taken in ~\citep{arcaute2015constructing, cottineau2017diverse} who identify cities by joining small sampling areas as long as they exceed a threshold value of population density. This approach is more flexible with respect to the type of data that we can use, but again requires somehow determining  appropriate threshold values.

We have a number of aims in this work:
\begin{enumerate}
    \item First, to develop a method for bounding cities and regions from any data, based on thresholds determined by the data itself.
    \item To apply this method to England and Wales as a case study. \begin{enumerate}
\item At the national scale we are interested in the influence and centralising effect of London, the UK's only mega-city. We will examine how the spatial distribution of wealth, jobs and infrastructure is more or less concentrated than population and how, say, an economic `lens' regionalises EW compared to a demographic or infrastructure one.
\item At the city level we are interested in finding city boundaries, studying when administrative or historic divisions are supported by the data. Again these boundaries can be different depending on the data and we study these effects for some chosen cities across a range of sizes.
    \end{enumerate}  
\end{enumerate}

\section{Methods}\label{sec:methods}

\subsection{Data}\label{sec:data}
We use data from the census of England and Wales reported by the UK's Office for National Statistics (ONS). We use data aggregated into Lower Layer Super Output Areas (LSOAs) which contain between 400 and 1200 households, with between 1000 to 3000 residents, since this is the finest resolution for which all of the data is available. We study the following types of spatial data
\begin{itemize}
    \item \textbf{Population}: This comes from (2021) census table TS001, which counts the number of usual residents in an LSOA.
    \item \textbf{Occupation}: This comes from census table TS062, NS-SeC classification.
    \item \textbf{Economic}: We use the ONS small area Gross Value Added (GVA) estimates \footnote{ \href{https://www.ons.gov.uk/economy/grossvalueaddedgva/datasets/uksmallareagvaestimates}{UK small area GVA estimates}}.
    \item \textbf{Transport}: We use the Department for Transport National Public Transport Access Nodes (NaPTAN) data \footnote{\href{https://www.data.gov.uk/dataset/ff93ffc1-6656-47d8-9155-85ea0b8f2251/naptan}{National Public Transport Access Nodes (NaPTAN)}} which contains data on the location of all public transport access points, such as bus stops and train stations.
\end{itemize}

Usual residents counts the number of people living in a household or communal establishment in each LSOA. The NS-SeC classification is the National Statistics Socio-economic Classification, which gives counts of individuals in various types of employment. We will focus on 
\begin{itemize}
    \item L1, L2 and L3 Higher managerial, administrative and professional occupations. Abbreviated L1.
    \item L7 Intermediate occupations.
    \item L10 and L11 Lower supervisory and technical occupations. Abbreviated L10.
\end{itemize}
See \href{https://www.ons.gov.uk/methodology/classificationsandstandards/otherclassifications/thenationalstatisticssocioeconomicclassificationnssecrebasedonsoc2010}{the ONS description} for more information. GVA is a regional equivalent of GDP, an estimate of the value generated by production of goods and services in an area. The NaPTAN data gives rail, metro and bus stop GPS locations, which are accumulated per grid square and LSOA.

The algorithm of \cite{arthur2024valeriepieris} for finding Valeriepieris circles requires gridded data. LSOA data is converted into a 0.01 square degree grid by assuming an even distribution across the LSOA and assigning to each grid square a fraction of the LSOA value equal to its overlap area with the grid square: 
$$v(g) = \sum_a \frac{ |LSOA_a \cap g|}{|LSOA_a|} v( LSOA_a )$$
where $g$ is a grid square, and $v(a)$ is the value of some statistic in a sampling polygon, $a$, with area $|a|$. LSOAs are, by design, quite variable in size, large in rural areas small in urban ones. To ensure that the final results do not depend on the arbitrary grid size, we tested a number of grid sizes and found extremely consistent results, since most of the regions we find are much larger than a single grid `pixel', see \ref{sec:appgrid}.

\subsection{Valeriepieris Circles}

The Valeriepieris(VP) circle was originally proposed by Ken Myers (under the pseudonym Valeriepieris) as a small circle containing over half the global population. The idea attracted attention from some academics \citep{quahtight,rami2018megacities} though mostly as a curiosity and always computed for the global population.  \cite{arthur2024valeriepieris} gave a more thorough treatment of the VP circle, defining it as the \emph{smallest} circle containing a fraction $f$ of the population of some area of interest. As noted in that work, the function 
$r(f)$, the radius of the VP circle computed for the population fraction $f$, is a transformation from two to one dimensions which still retains important information about the two dimensional data distribution.

If the population density at coordinate $\vec{x}$ is $\rho(\vec{x})$ then the population in a disc $D(\vec{c}, r)$ with centre $\vec{c}$ and radius $r$ is
\[
p(D(\vec{c}, r)) = \int_{D(\vec{c}, r)} d\vec{x} \rho (\vec{x})
\]
The VP $f$-circle defines the disc of smallest radius where
\[
p(D(\vec{c}_{VP}(f), r_{VP}(f))) = fP 
\]
where $P$ is the total population. If the population was distributed isotropically around $\vec{c}_{VP}$ we could set that point as the origin and then
\[
fP = 2\pi \int_0^{r_{VP}(f)} r dr \rho (r)
\]
We will hereafter drop the VP subscript for clarity, assuming we are always discussing the Valeriepieris, i.e. smallest, circle.

Previous work has assumed an exponential decay for the population density as a function of distance from a city's central business district \citep{clark1951urban, makse1998modeling}
$$\rho(r; b) \propto \exp\left( -br \right)$$
In \cite{arthur2024valeriepieris} a power law form was used 
$$\rho(r; a) \propto r^{a-2}$$
see also \cite[Chapter~8]{batty1994fractal} while in \cite{guerois2008built} a fixed value of $a$ was used ($a=1$). Other more complex models, like S-curves \citep{jiao2015urban} and quadratic gamma type functions \citep{zielinski1979experimental} have also been considered but we will show that the simple power law is an excellent fit to the data.

A dense core city, like London, is likely to contain the smallest circle with a sizable fraction of the population. In a densely populated city, this circle only needs to be slightly 
expanded to increase the included population fraction substantially. As we move out from the core, the density decreases and the circle will have to expand more to increase the included 
fraction by the same amount. The city will at some point turn into sparsely populated countryside at which point we switch to a different regime, where the (low) density changes at a slower 
rate. Thus a change in the \emph{rate} of density decay, rather than in density itself, is indicative of an urban-to-rural transition. This is reflected by a change in the rate of change of 
$r(f)$ with $f$. As the circle expands through the countryside, it gradually meets other large cities, far enough away not to have been `absorbed' by the core city. When these cities are 
reached, the growth rate of the circle slows again, reflected by another abrupt change in the rate.

Motivated by this, we will use a model where the population is distributed in a series of concentric rings. The population density will follow the same parametric form across all rings, though with different parameters within each ring. With a simple one-parameter density we have
\begin{align*}
\rho(r) = \begin{cases}
c_1\rho(r; \alpha_1) \qquad 0 < r < r_1\\
c_2\rho(r; \alpha_2) \qquad r_1 < r < r_2\\
\ldots\\
c_k\rho(r; \alpha_k) \qquad r_{k-1} < r < r_k\\
\end{cases}
\end{align*}
where $c_i$ are chosen to make the density continuous at the boundaries. 

If we have to go to a distance $r(f)$, taking us into the $j^{th}$ ring, to capture a fraction $f$ of the population then
\[
fP = C + 2 \pi c_j\int_{r_{j-1}}^{r(f)} r dr \rho (r) = C + 2 \pi c_j( \hat{ \rho } (r(f)) - \hat{ \rho } (r_{j-1}))
\]
where $\hat{ \rho } = \int r dr \rho(r)$ and $C$ is the population up to the $(j-1)^{th}$ ring. For a power law parametrisation the integral is elementary and, within a single ring, there is a linear relationship between $f$ and $r_f$ on a log-log plot. Combining all of the rings gives a piecewise linear relationship between $r_f$ and $f$ as $f$ ranges from 0 to 1. We use the power law density throughout, we have tried the exponential form and found the fit to be much less convincing and the breakpoints harder to determine.

This model has been shown to work well for the UK and other countries in \cite{arthur2024valeriepieris}.  A similar method was applied in the work of \cite{guerois2008built} 
(see also \cite{wu2021characterizing}), who use an expanding circle with origin at the historic city centre to produce similar sorts of profiles with CORINE land cover data by 
averaging over these circles. \cite{taubenbock2019new} also use a radial model and seek transition points to identify city boundaries. The advantage of our method is that the 
centres can be determined from data and can even move as the included fraction changes. The VP radius considered as a function of $f$ is a transformation, mapping the population, 
or any other, density $\rho(\vec{x})$ to a ring model $\rho(r)$ in such a way that the populations and radii of the VP $f$-circles for the ring model and real data are the same. 
This `ring city' is much easier to study than the original one while retaining many of its important characteristics. In particular the profile, $r(f)$, is much simpler than the ones found by \cite{guerois2008built} and \cite{taubenbock2019new} and depends only on the data rather than  historic definitions of the centre. It also accounts for situations when, say, the historic centre and current business centre are in different places. This automatic centering means it can also be applied to larger areas, like whole regions or conurbations of multiple cities, where there is no obvious fixed centre point. 

In the ring model it is natural to examine the breakpoints which separate different regimes of population density dependence. We fit a piecewise linear curve to the log transformed data using the \emph{pwlf} library \citep{pwlf}. The optimal number of breakpoints is determined by hand. Model selection methods, like the Bayesian or Akaiake Information Criterion, struggle since the radii are strongly correlated - the radius for the fraction $f$ is going to be very similar to the radius for the fraction $f + \delta f$ - and therefore standard estimates of the model likelihood will underestimate the error. Fortunately, it is rather obvious how many breakpoints to choose in all the examples we consider though an automatic method to detect the number of breakpoints would be useful to develop in future.

\subsection{City Boundaries}

We apply essentially the method of \cite{arcaute2015constructing}, clustering adjacent sample areas if they exceed some density threshold $\rho_0$. Areas with lower density but which are surrounded by cells above the threshold (e.g. city parks) are also included so the final area has no holes. We will determine the density threshold as the value of density at the boundary between rings. The piecewise linear fit gives the parameters of the density function in each ring. Using the power-law parametrisation, in the $j^{th}$ ring
\begin{align*}
    %\log (fP) &= \log (C- c_j \hat{ \rho } (r_{j-1})) + \log\left( \frac{c_j r^{a_j}}{a_j}\right)  \\
    \log(r) &= \frac{1}{a_j} \log(f) + C
\end{align*}
for some constant $C$, so the parameter $a_j$ is given by reciprocal of the slope of the $j^{th}$ linear segment fit to the data. Imposing continuity of $\rho$ fixes all but one of the $c_j$ via
\begin{align*}
   c_{j+1} &= c_j \frac{ r^{a_j-2}}{r^{a_{j+1}-2}}
\end{align*}
The one remaining constant is fixed by requiring that the integral of $\rho$ over the entire area has to give the total population $P$,
\begin{align*}
   P = 2 \pi \sum_j \frac{c_j}{a_j} (r_j^{a_j}-r_{j-1}^{a_j} )
\end{align*}
which can be solved for the one remaining free $c_j$. 

To determine a city boundary we start with the rough location of a city of interest and a bounding box containing the city and its hinterland, similar to other boundary finding methods reviewed in \cite{montero2021delineation}. To focus on the city of interest when there are multiple cities in close proximity we only search within a given distance of the approximate centre. Typical box sides lengths are 30km (50km for London) and 5km for the search radius. The final results are not extremely sensitive to these numbers, as long as they are `reasonable', see \ref{sec:appgrid}. We compute the VP profile for this region and identify the boundary from the last breakpoint which, in the ring model, is the point at which the regime changes from city to country. The threshold density is then 
\begin{equation}
\rho_0 = \frac{c_b}{a_b} r_b^{a_b-2}
\end{equation}
and all grid cells with density greater than this are merged. We will call this the Valeriepieris (VP) boundary. We compare the city boundaries found from this process with the Office 
for National Statistics Major Towns and Cities boundary polygons 
\footnote{ \href{https://geoportal.statistics.gov.uk/datasets/ons::major-towns-and-cities-december-2015-boundaries-ew-bgg-v2/about}{ONS major towns and cities}
} where available.

\begin{figure}[h!]
    \centering
    \includegraphics[height=0.85\textheight]{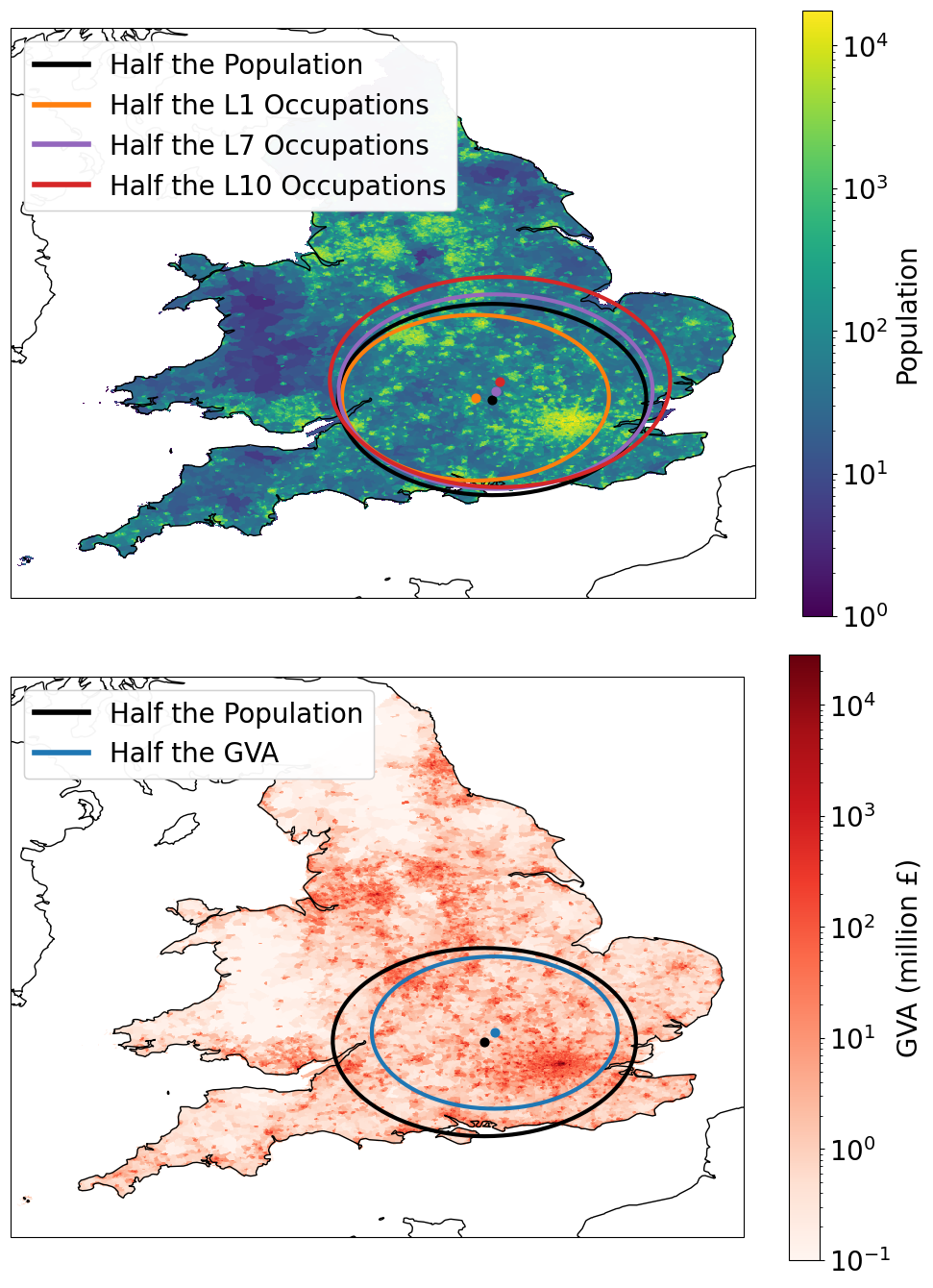}
    \caption{Top: population, Bottom: GVA. Circles are the 50\% Valeriepieris circles (smallest circles containing half the data). The population circles are shown in both for reference.}
    \label{fig:fig1}
\end{figure}
\begin{figure}[h!]
    \centering
    \includegraphics[height=0.425\textheight]{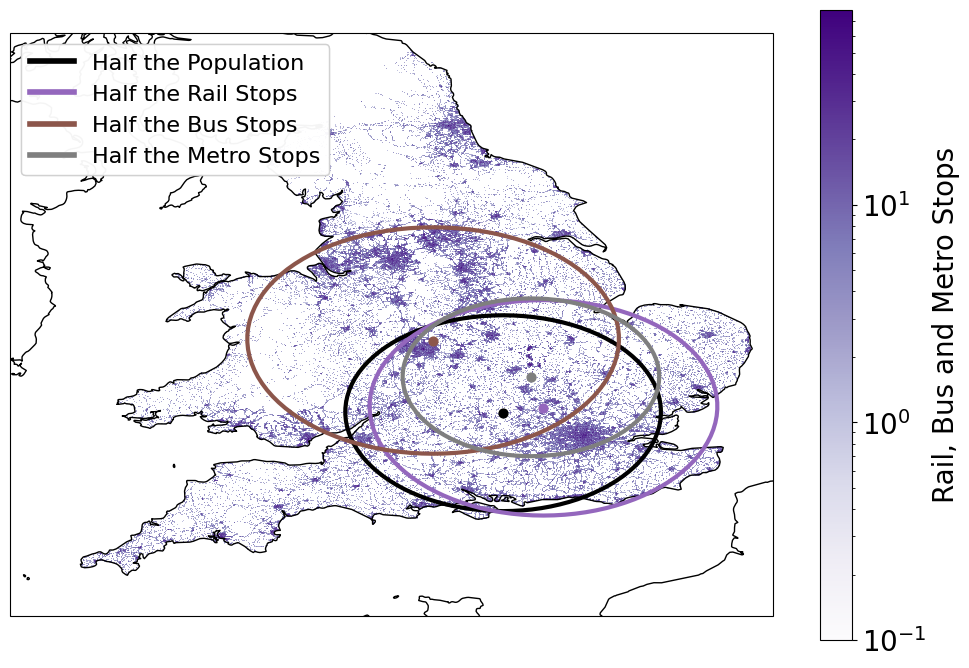}
    \caption{ NaPTAN Rail, Metro and Bus stops. The vast majority of the data are bus stops. The circles are the 50\% Valeriepieris circles (smallest circles containing half the data). The population circle is shown for reference.}
    \label{fig:fig2}
\end{figure}

\section{Results}\label{sec:results}

\subsection{Valeriepieris Circles for England and Wales}

\begin{table}[h]
    \centering
    \begin{tabular}{|c|c|c|}
    \hline
      & VP centre & VP radius (km)\\ \hline \hline
      Population   &  (51.765, -1.005) & 121.4 \\ \hline \hline
      GVA   & (51.875, -0.885)& 98.2 \\ \hline \hline
      L1 occupations & (51.785, -1.195)& 105.1 \\ \hline
      L7 occupations & (51.855, -0.965)& 123.4 \\ \hline
      L10 occupations & (51.965, -0.915)& 133.5 \\ \hline \hline
      Rail Stops & (51.825, -0.5549) & 133.6 \\ \hline
      Metro Stops & (52.165, -0.695) & 97.8 \\ \hline
      Bus Stops & (52.575, -1.7850) & 140.4 \\ \hline
    \end{tabular}
    \caption{50\% Valeriepieris centres and radii for the datasets. }
    \label{tab:tab1}
\end{table}

Figure \ref{fig:fig1} shows the data for GVA and number of individuals in L1,L7 and L10 occupations with the VP circles containing 50\% of the GVA/occupations compared to the VP circle containing 50\% of the population. Table \ref{tab:tab1} shows the values of the centres and radii. The centres are all quite close, north west of London, not too far from the city of Oxford. The population circle is large enough to capture Greater London and Birmingham. In these and subsequent tables and figures, population is used as a baseline or null model. If the boundaries for some other data are different, this tells us if that variable is more or less concentrated than population.

 The GVA and L1 circles are slightly smaller than the population circle. The GVA circle misses some cities south and west of London and does not completely cover Birmingham. The L1 circle is shifted west, excluding East Anglia (rural) and southern satellite towns and cities, it is just large enough to capture Gatwick airport. The L7 circle is quite close to the population circle and the L10 circle is larger, capturing more of the population of the midlands and north. These circles show firstly, the demographic influence of London and the south east,  most of the people in EW live there. The GVA and occupation circles show that the economic influence of London and the south east is greater than its population alone would imply. Wealth and high prestige L1 jobs are more concentrated and conversely the, likely lower pay, L10 occupations are sparser in the south east.

Figure \ref{fig:fig2} shows the NaPTAN data for Rail, Metro and Bus stops. Note that there are only around 1100 metro stops and 2700 rail stops compared to 370,000 bus stops, so bus stops would completely overwhelm a combined dataset and each transport mode is therefore treated separately. As Figure \ref{fig:fig2} and Table \ref{tab:tab1} show, the rail and metro circles are relatively similar to the population one, located in the south east. The metro circle is quite small, likely due to London's extensive underground system and the rail circle is larger than population one, perhaps reflecting the fact that national rail into London is focussed on a few large terminals and onward travel in the city often proceeds via the metro system. 

The bus circle is notably different. The centre is in Birmingham and it does not even include all of London. We will see later that smaller fractions than $f=0.5$ do centre the VP circle on London, but this plot is an indication that the bus network (which is a function of the road network) does not necessarily reflect the same regionalisation as population or economic data and also raises interesting questions around possible inequalities in accessibility (by bus transport) of jobs and services in different parts of the country.

\subsection{Population Boundaries}\label{sec:popprof}

\begin{figure}[h!]
    \centering
    \includegraphics[height=0.9\textheight]{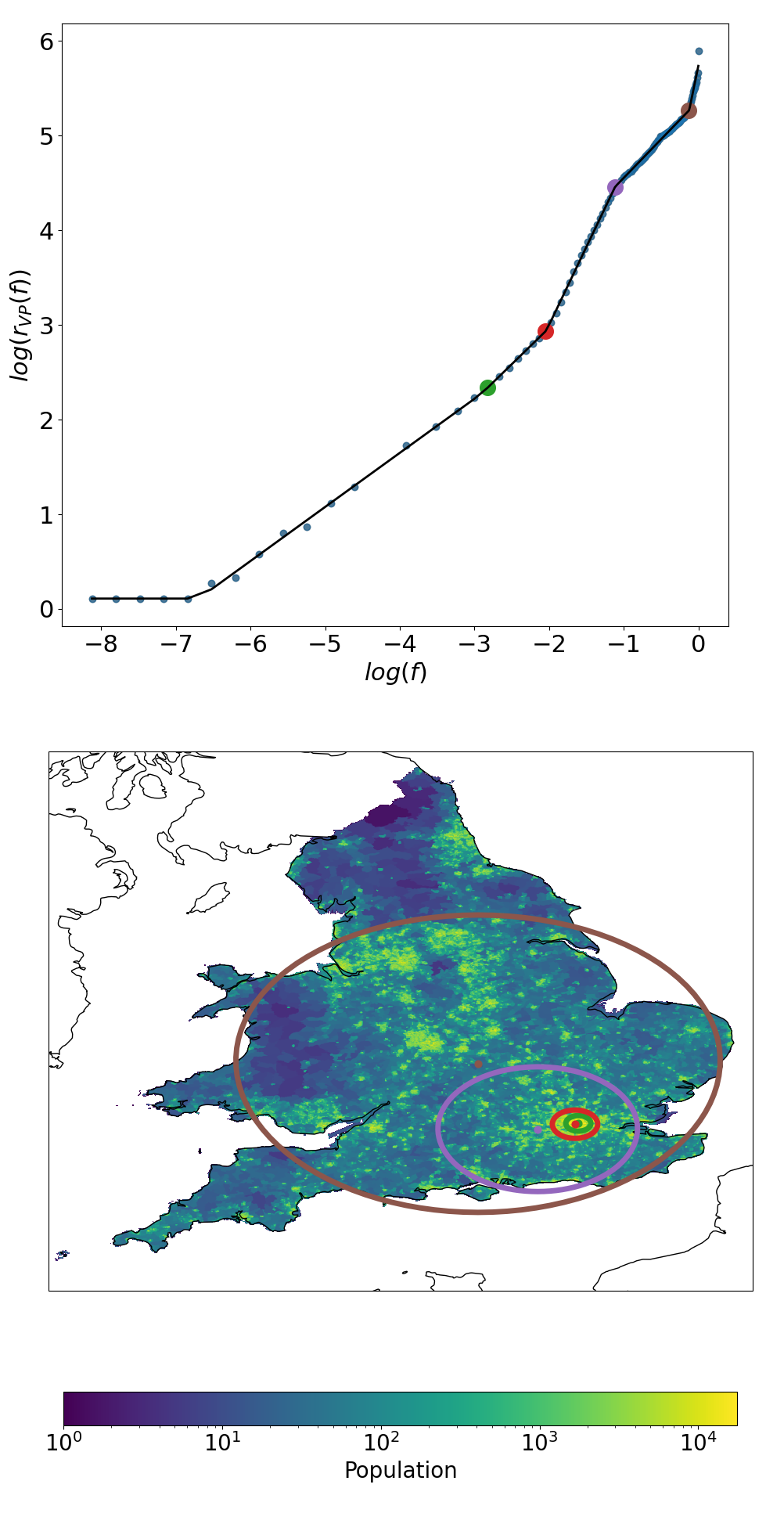}
    \caption{Top: VP-profile with breakpoints highlighted. Bottom: England and Wales population with VP circles shown for each breakpoint.}
    \label{fig:popbound}
\end{figure}
\begin{table}[h!]
    \centering
    \caption*{Population Boundaries}    
    \begin{tabular}{|c|c|}
    \hline
        $f$ & $r(f)$ (km) \\ \hline \hline
0.055 &  10.5\\ \hline
0.131  & 19.0\\ \hline
0.328 &  83.7\\ \hline
0.885 &   199.5\\ \hline
    \end{tabular}
    \caption{Fraction of included population at the breakpoints and the corresponding VP radii. }
    \label{tab:tab2}
\end{table}

The VP-profile for population data is shown in the top panel of Figure \ref{fig:popbound}. The fit is performed with 5 breakpoints with their location  determined by minimising the residual sum of squares using the \emph{pwlf} library. The first breakpoint (between -6 and -8) is due to grid artefacts and is not informative. The others are highlighted in the figure and shown in Table \ref{tab:tab2}. The first break, at the green circle, corresponds to `inner London'. The next break, at the red circle, encompasses most of greater London, followed by the purple circle which encompasses the south east. The brown circle, which contains most of the population, encompasses the large cities of the north and midlands.

There is some suggestion of a bump in the profile between the south east and the rest, which would put the midlands in a separate zone from the north. However the change in slope is quite small if we fit with an additional breakpoint, so we include the Midlands with the North as there is not strong support from the data for their separation. The circles define some fairly natural boundaries for England and Wales. London is at the core, then the south east, then a much larger radius encompasses most of the rest of the population in a zone outside London. Perhaps most interesting are the places outside this outer circle. These include western Wales, Devon and Cornwall and the North East. These are what we could reasonably call the periphery of EW and are places with distinct historical influences e.g. Scandanvian influences in the north east and Celtic in the south west and Wales, the latter even have persistent minority languages, despite centuries of assimilation.
\begin{figure}[h!]
    \centering
    \includegraphics[width=0.5\textwidth]{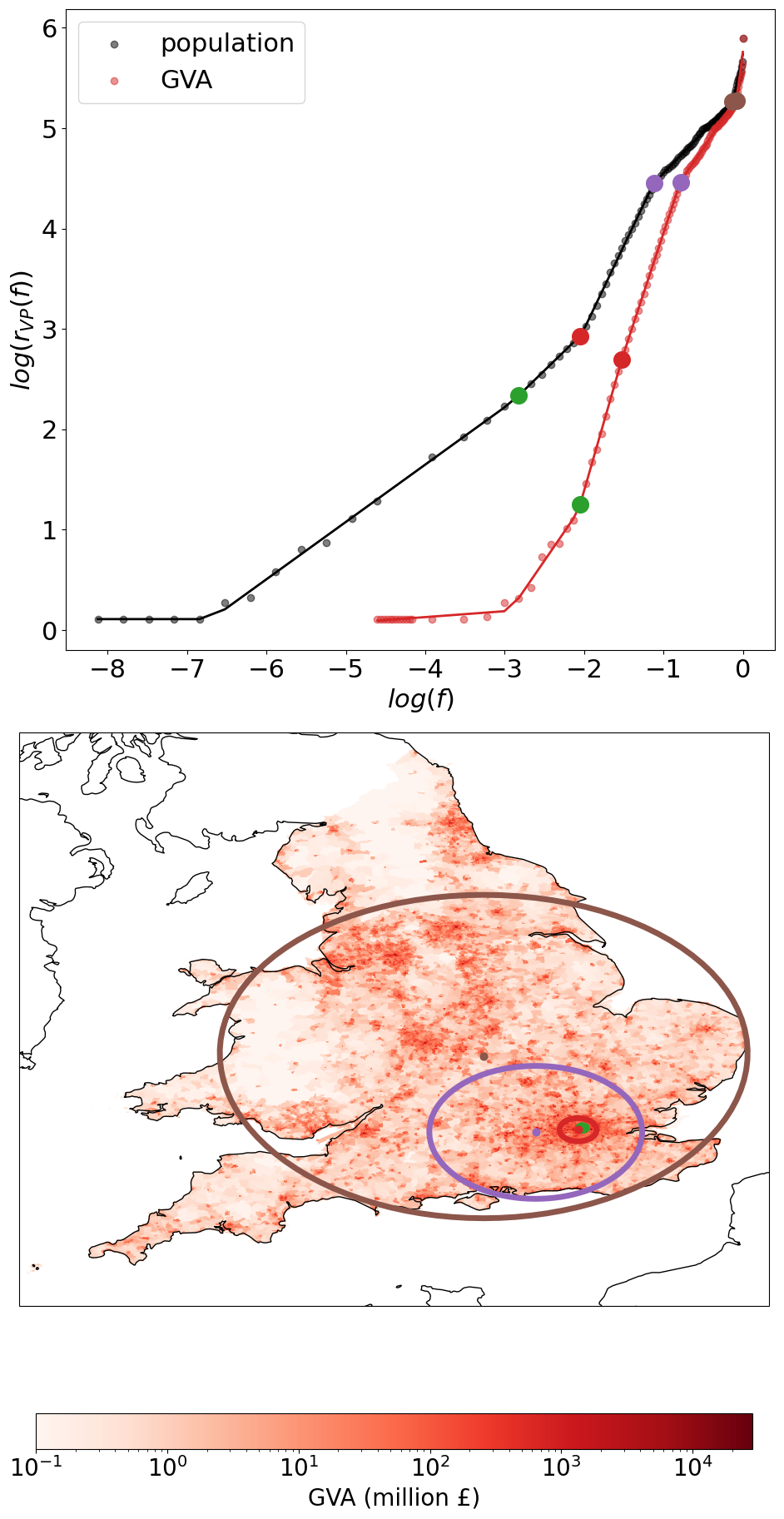}
    \caption{ Top: VP-profiles for population and GVA with breakpoints highlighted. Bottom: Map showing the GVA data with the boundaries derived from the breakpoints above.}
    \label{fig:gvabound}
\end{figure}
\newpage
\subsection{England and Wales GVA Boundaries}

\begin{table}[h]
    \centering
\caption*{GVA Boundaries}    
    \begin{tabular}{|c|c|}
    \hline
$f$ & $r(f)$ (km) \\ \hline \hline 
0.125 & 3.5 \\ \hline
0.222 & 14.6 \\ \hline
0.458 & 83.9 \\ \hline
0.924 & 204.0 \\ \hline
    \end{tabular}
    \caption{Fraction of included data at the breakpoints and the corresponding VP radii. }
    \label{tab:tab4}
\end{table}
Figure \ref{fig:gvabound} and Table \ref{tab:tab4} shows the boundaries derived from the GVA data. Interestingly, the radii and centres are very similar, suggesting the zones found from the population data Figure \ref{fig:popbound} are robust to changes in exact data used to determine them. There are however, a few major differences. First is the fraction of the data included at at each boundary. Most strikingly, a roughly 3.5km zone in central London contains around 13\% of the total GVA. In the population data the first boundary occurs a little further out, 10.5 km, and only contains 5\% of the population. This highlights that central London is an outlier, adding a disproportionate amount to the total GVA. Outside of central London, the boundaries are largely the same as found for population, consisting of Greater London, the South East and the rest.

\begin{figure}[h!]
    \centering
    \includegraphics[width=0.95\textwidth]{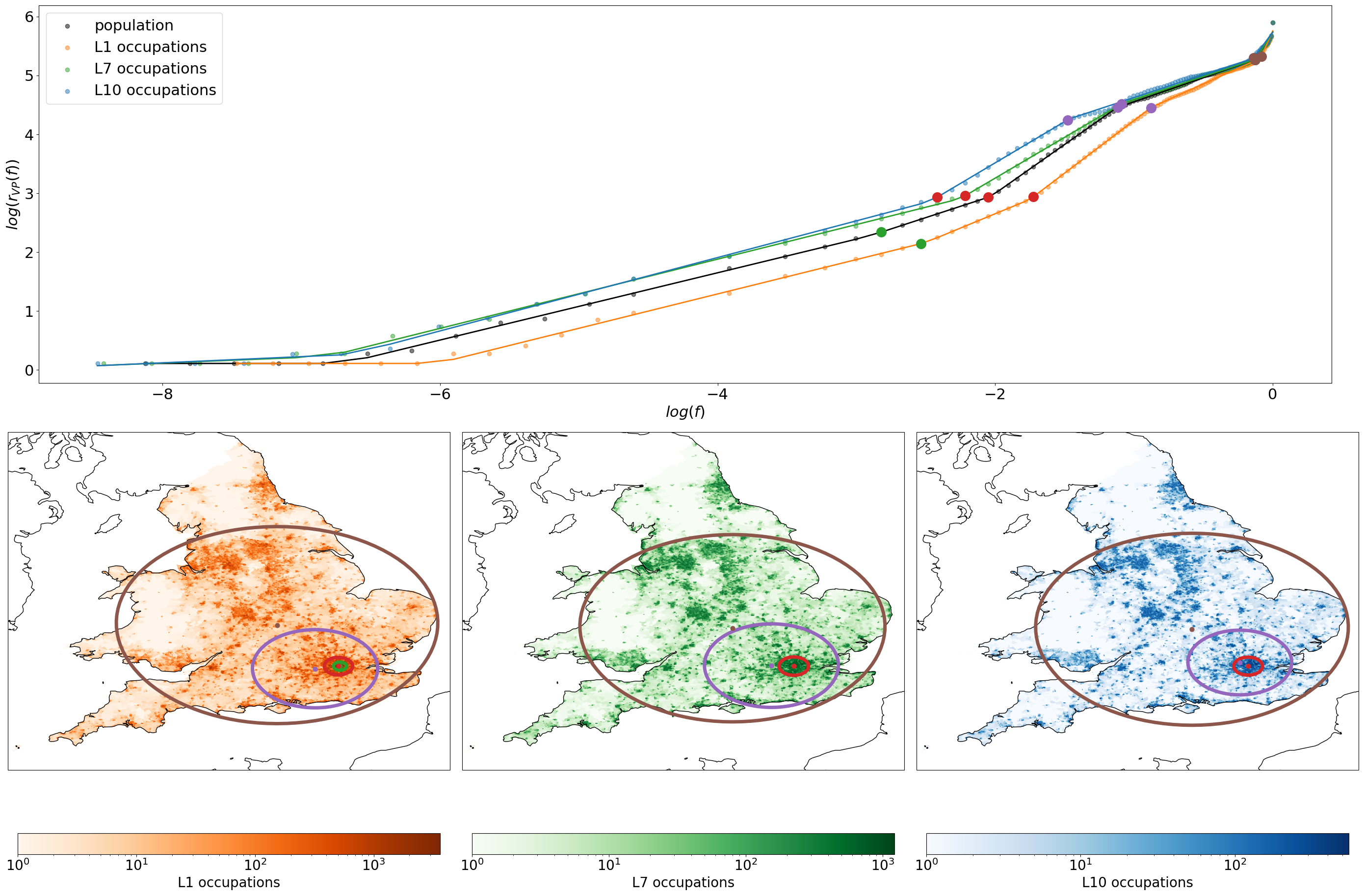}
    \caption{ Top: VP-profiles for population and occupation data with breakpoints highlighted. Bottom: Maps showing the occupation data with the boundaries derived from the breakpoints above.}
    \label{fig:occbound}
\end{figure}
\newpage
\subsection{England and Wales Occupation Boundaries}

\begin{table}[h]
    \centering
\caption*{Occupation Boundaries}    
    \begin{tabular}{|c|c|c|}
    \hline
Data & $f$ & $r(f)$ (km)\\ \hline \hline 
L1 & 0.082 & 8.7 \\ \hline
\hline
L1 & 0.174 & 18.9 \\ \hline
L7 & 0.105 & 19.6 \\ \hline
L10 & 0.085 & 19.2 \\ \hline
\hline
L1 & 0.418 & 83.7 \\ \hline
L7 & 0.340 & 89.5 \\ \hline
L10 & 0.231 & 69.2 \\ \hline
\hline
L1 & 0.916 & 211.0 \\ \hline
L7 & 0.884 & 200.5 \\ \hline
L10 & 0.868 & 205.6 \\ \hline
    \end{tabular}
    \caption{Fraction of included data at the breakpoints and the corresponding VP radii. }
    \label{tab:tab5}
\end{table}

Figure \ref{fig:occbound} and Table \ref{tab:tab5} show the boundaries derived from the occupation data. Like GVA, but less pronounced, L1 occupations are concentrated in inner London, with around 8\% of these jobs located in an 8.7km zone in central London. For L7 and L10 occupations in contrast, there is not even any support in the data (change of slope in $r(f)$) for separating inner and outer London. Outside of inner London, for these categories of occupations the boundaries found are more or less similar to population. Indeed the L7 and L10 profiles are shifted left of the population profile - indicating that fewer of these jobs are contained within the boundaries of London and the south east. So London and the south east have more prestigious jobs than their population would suggest and conversely the midlands and north have more `ordinary' jobs.

\begin{figure}[h!]
    \centering
    \includegraphics[width=0.5\textwidth]{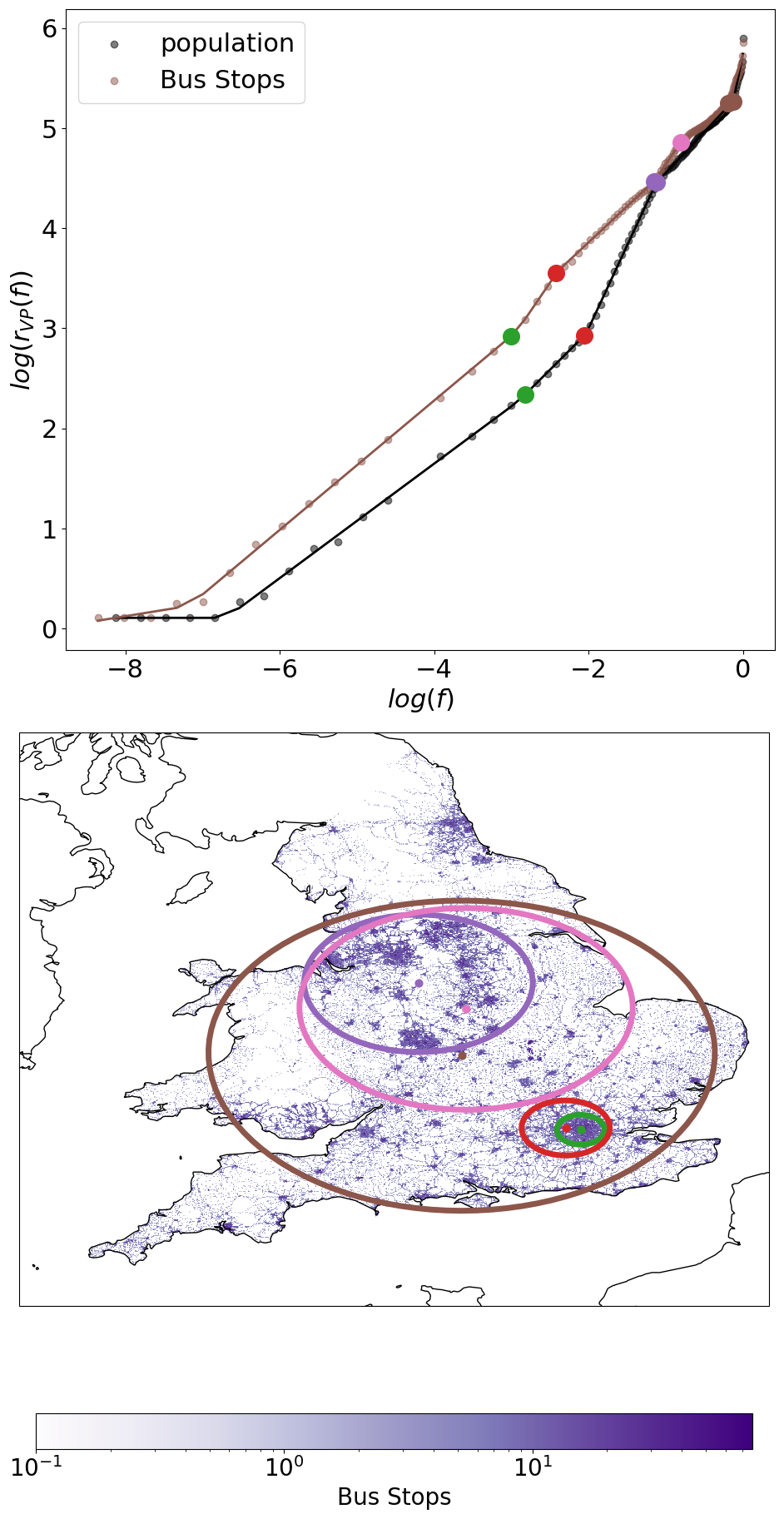}
    \caption{ Top: VP-profiles for population and bus stops with breakpoints highlighted. Bottom: Map showing the bus stop data with the boundaries derived from the breakpoints above.}
    \label{fig:busbound}
\end{figure}
\begin{table}[h]
    \centering
\caption*{Bus Stop Boundaries}    
    \begin{tabular}{|c|c|}
    \hline
$f$ & $r(f)$ (km)\\ \hline \hline 
0.053 & 18.7 \\ \hline
0.091 & 34.7 \\ \hline
0.320 & 86.7 \\ \hline
0.443 & 127.3 \\ \hline
0.828 & 195.7 \\ \hline
    \end{tabular}
    \caption{Fraction of included data at the breakpoints and the corresponding VP radii. }
    \label{tab:tab6}
\end{table}

\newpage
\subsection{England and Wales Transport Boundaries}

Figure \ref{fig:busbound} and Table \ref{tab:tab6} show the boundaries determined from the bus stop data. The rail and metro data is too sparse for this method to be effective. Figure \ref{fig:busbound} has a number of striking differences compared to the other data. While the first two boundaries in inner and outer London are similar to other data sets, once London is covered the centre of the minimum circle moves dramatically north west. So while London has this highest bus stop density, bus stops in the south east are relatively sparse. Thus the smallest circle containing larger fractions of the bus stops is actually in the north west, encompassing the large northern cities and conurbations. Also of note is the extra boundary (pink) due to the `gap' between the dense bus network in the north west and greater London. The outer boundary is as before, with the same peripheral places identified. Finally note that despite the large shift in the physical location of the VP centre, the profile in Figure \ref{fig:busbound} is still smooth and fit very well by the ring model.

\begin{figure}[h!]
    \centering
    \includegraphics[width=0.75\textwidth]{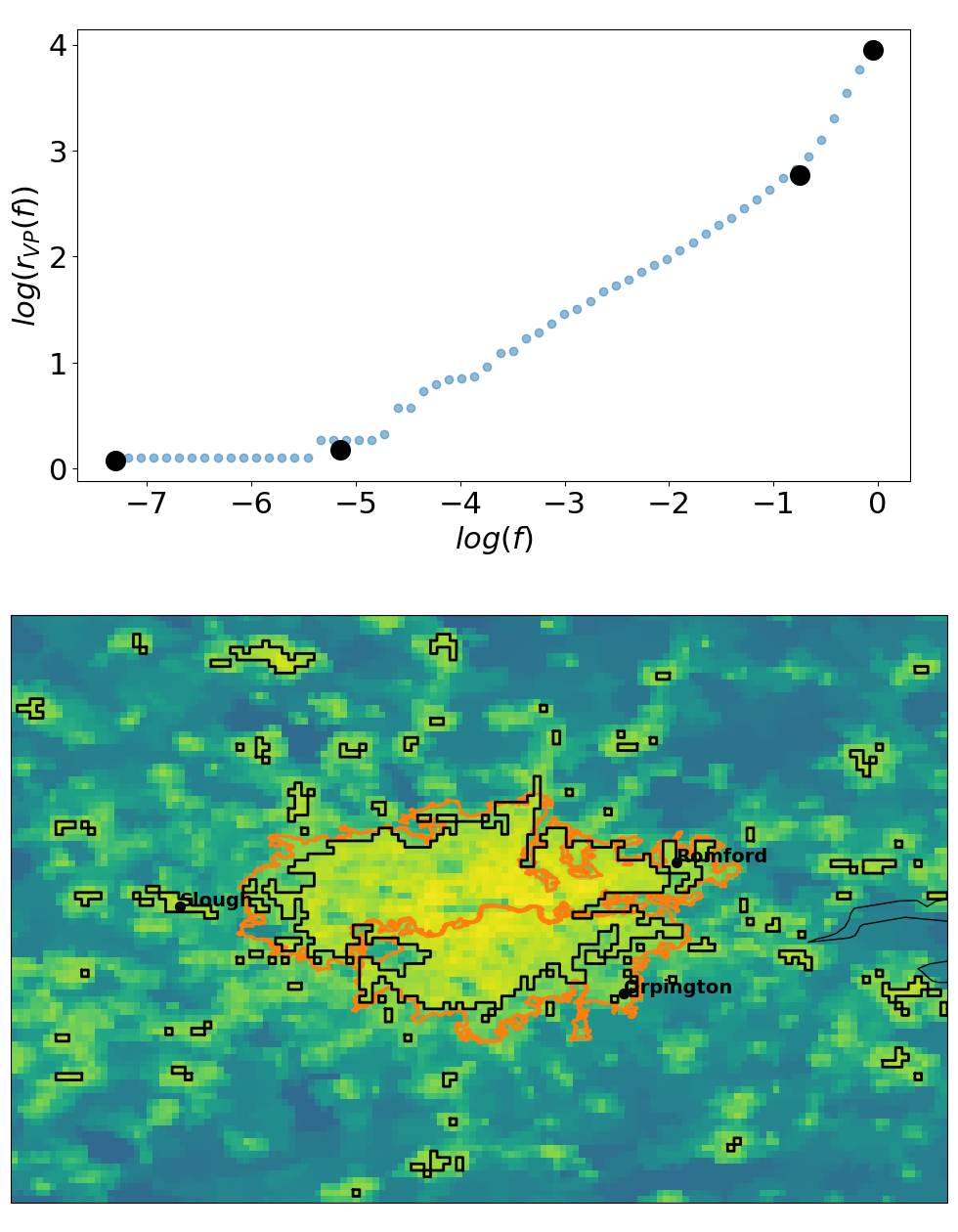}
    \caption{ Top: VP profile and breakpoints for a 50km box centred at (51.507222, -0.1275). Bottom: VP Boundary (black)  and the official boundary (orange) shown with population data. The VP boundary is determined from grid squares with density greater than 3851 people per km$^2$.}
    \label{fig:londonpop}
\end{figure}
\subsection{City boundaries from Population}

A boundary for London, determined as described in Section \ref{sec:methods}, is shown in Figure \ref{fig:londonpop}. The largest contiguous area could be taken as the city boundary, but it is also interesting to show nearby areas which also exceed the density threshold. The largest polygon overlaps quite well with the official boundary used by the Office for National Statistics but some differences are notable - near the suburbs of Orpington and Romford, and a handful of other small `islands', the population dips below the density threshold. We could think of these places as separate towns rather than part of London's outskirts since there is a significant dip in density when traveling from London to, say, Orpington.

\begin{figure}[h!]
    \centering
    \includegraphics[width=0.75\textwidth]{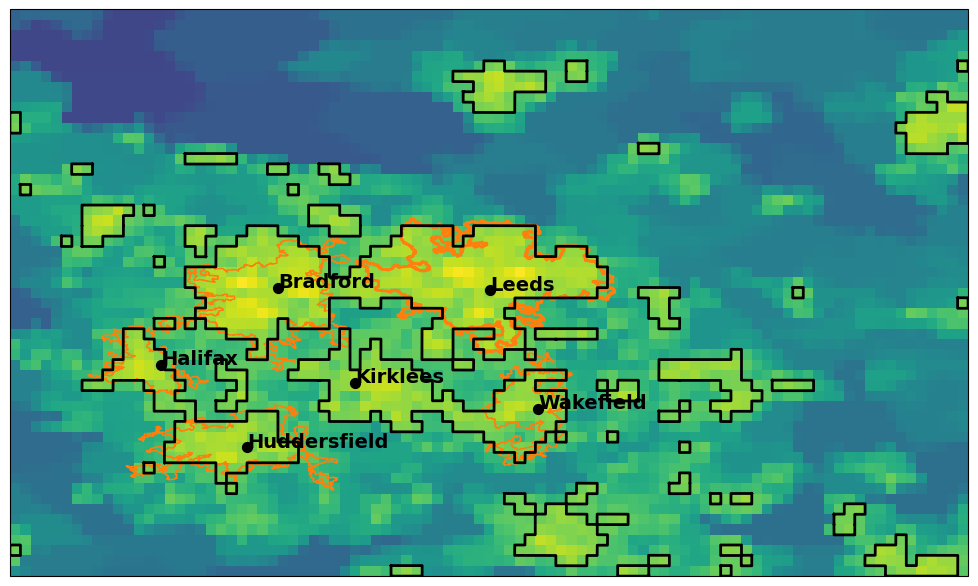}
    \caption{ VP Boundary (black) and the official boundary (orange) shown with population data for Leeds-Bradford. The critical density is determined using a 30km box centred at (53.7975, -1.543611) which suggests including grid squares with over than 1214 people per km$^2$.}
    \label{fig:leedspop}
\end{figure}

Figure \ref{fig:leedspop} shows the Leeds-Bradford area. According to ONS boundaries, the towns and cities in this conurbation are still distinct. This analysis shows however that there is sufficient density between Leeds and Bradford to consider them as one place. Leeds suburbs become Bradford suburbs without any notable thinning of population. This is also the case for Wakefield and the area of Kirklees (which is not an ONS designated Major Town or city) and Halifax and Huddersfield. These three city pairs are themselves very close to overlapping, and indeed using a different grid or (as we will see in the next section) different data can make the cities merge. We take this as an indication that this area is on the precipice of becoming one large conurbation or polycentric city. 

\begin{figure}[h!]
    \centering
    \includegraphics[width=0.7\textwidth]{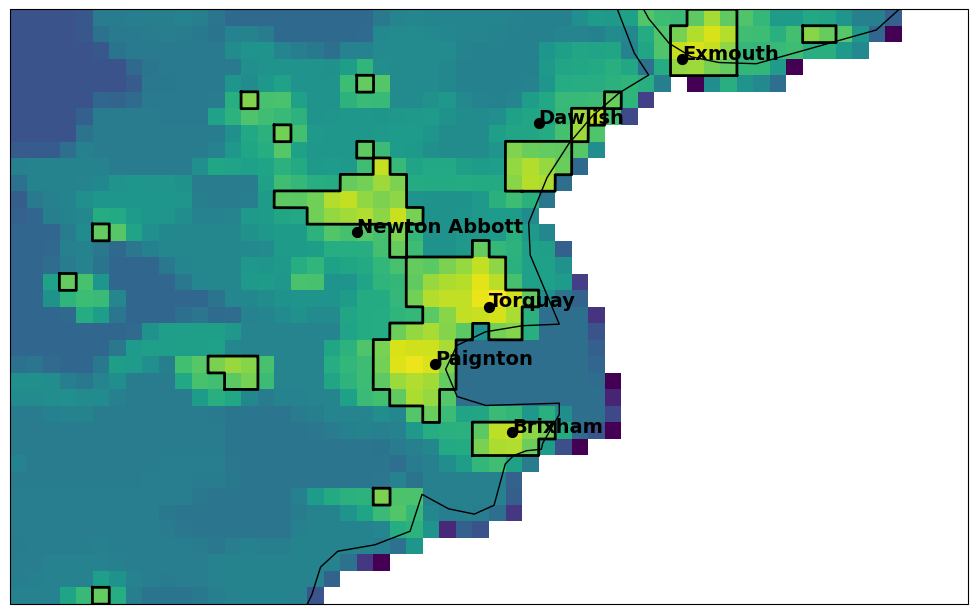}
    \caption{ VP Boundary (black) shown with population data for Torbay area. The critical density is determined using a 20km box centred at (50.47, -3.53) which suggests including grid squares with over than 703 people per km$^2$.}
    \label{fig:torquay}
\end{figure}

Figure \ref{fig:torquay} shows a similar story as for Leeds-Bradford on a smaller scale for the Torbay area. The historically distinct towns of Torquay and Paignton have merged while the nearby towns of Newton Abbott and Brixham are still separated by low density gaps, with Newton Abbott very close to merging with the Torbay area. 

\subsection{City boundaries from Other Data}

\begin{figure}[h!]
    \centering
    \includegraphics[height=0.85\textheight]{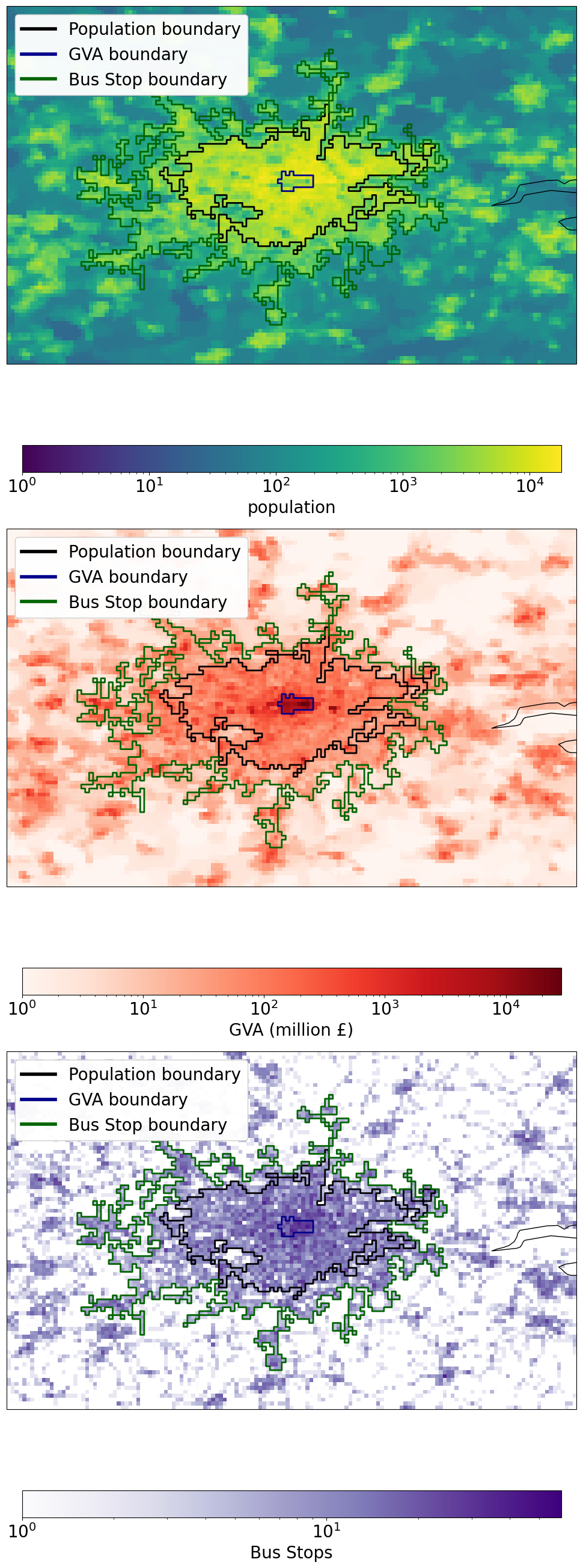}
    \caption{ VP Boundaries determined from population (black), GVA (blue) and bus stops (green) for London.}
    \label{fig:londonalt}
\end{figure}
\begin{figure}[h!]
    \centering
    \includegraphics[height=0.85\textheight]{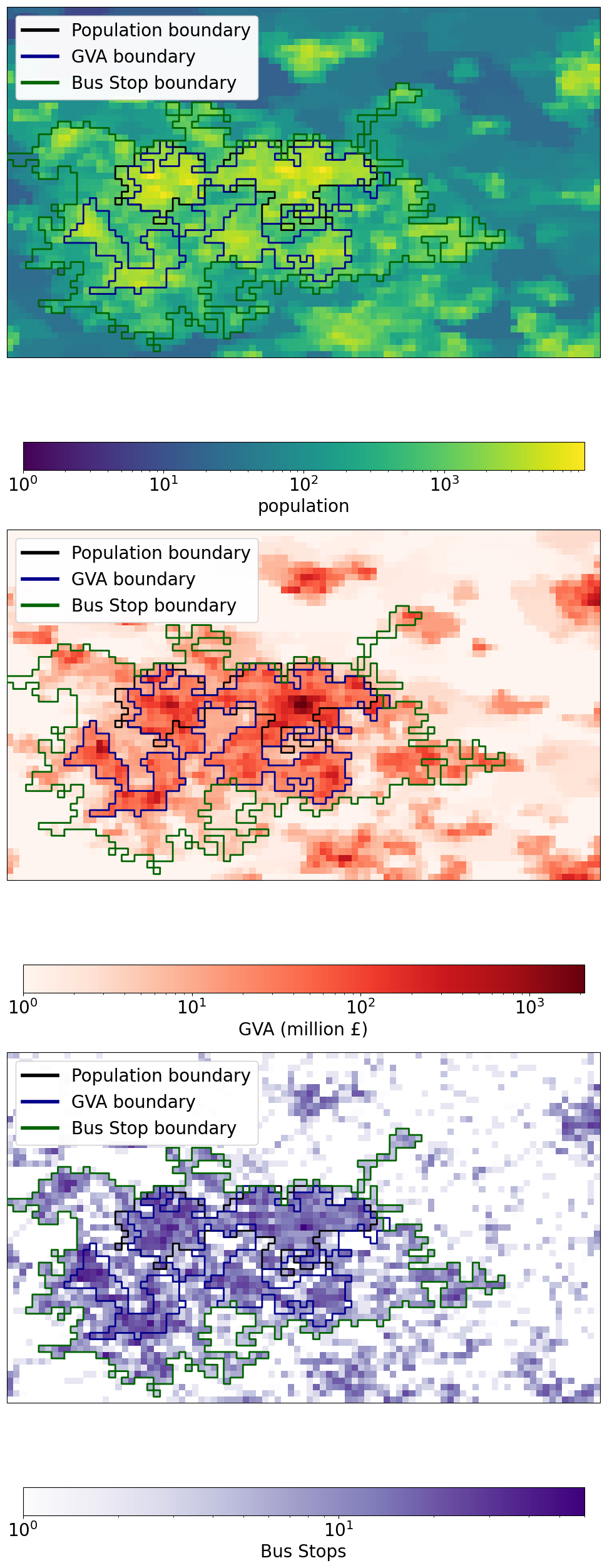}
    \caption{ VP Boundaries determined from population (black), GVA (blue) and bus stops (green) for Leeds-Bradford.}
    \label{fig:leedsalt}
\end{figure}

Figure \ref{fig:londonalt} shows London and Figure \ref{fig:leedsalt} shows Leeds, but this time computing the boundary using GVA and bus stops as well as population. For London, the GVA boundary is much smaller than the population derived one and encompasses only inner London. This indicates firstly, that the economic output measured by GVA is much different in inner London than in the rest of the city and secondly that outer London's economic activity is not distinct enough from the surrounding towns to suggest another boundary within the 50km sampling area. As shown in Figure \ref{fig:gvabound}, if a larger area in the south east was used we would see such a transition. The boundary determined from bus stops is larger, as might be expected, incorporating more of the peripheral towns inside, where the density of communting links remains high even if population and GVA density do not. This demonstrates that city boundaries identified using different data sets can differ, which raises interesting questions on what the most appropriate way of identifying cities is.

Leeds-Bradford is more typical of most cities in EW, in that the GVA data is not as different from population data as is the case in London. In contrast to the population data, the GVA data does merge all of the historically distinct towns of this area while the population data keeps them (just about) separate. As discussed for the population data, this region seems on the verge of merging into one large urban area, with GVA data suggesting this has already occurred and population data suggesting they have not quite merged. The bus boundary encompasses the population and economic boundaries and indicates a dense bus network within this multi-city urban area.

\section{Conclusion}

Our analysis of England and Wales confirms well known social facts \citep{mccann2020perceptions} from simple count data - the centralising effect of London, the North South divide, the peripheral status of areas in the North East and South West. Regional boundaries are robust to the underlying data used to derive them, with the similar boundaries discovered using population, GVA and occupations. A notable exception is bus stop data, which corresponds to a kind of backbone for the road network. In this case, perhaps due to the more extensive rail and metro services in the south east, the bus data suggests separate and sparsely connected north-west and south-east zones.

City boundaries found from population data map closely to known administrative boundaries, with the same methods applying to large and small cities (London $\sim 8.5$ million residents, 
Torbay $\sim 100$ thousand residents). Using a density threshold derived from the data shows when areas have effectively merged to become polycentric \citep{cai2017using} cities 
(Leeds-Bradford) or could still be considered separate (London/Slough). Using different data can often result in different city boundaries, for example in London the large amount 
of wealth concentrated in the inner city makes this distinct from the outer part of greater London which, when it comes to GVA, is not distinct from nearby towns and cities. 
Understanding the many different `Londons' is an interesting research programme \citep{paul2017limits, vanhoof2019using}. Our methods show that different data can reveal different 
faces of a city, and suggest that a single, universal definition of \emph{the city} is difficult to achieve, as city boundaries differ depending on what data set is used to identify them.

The methodology we have developed based on Valeriepieris circles and their growth is quite general. We have applied it to various different data sets and found it to be robust and to return known patterns in population and wealth distribution. There is nothing specific to England and Wales, apart from easily accessible census data, required for this analysis. Indeed the density and proximity of many of its cities makes finding their boundaries much more challenging than in countries like the US, which tends to have better defined and separated cities. The transformation and radial disc model is quite effective at identifying breakpoints in the data that have real social meaning for the data set under investigation. This can help address the so-called `Zonation Effect' \citep{manley2021scale}, an aspect of the Modifiable Aeral Unit Problem (MAUP), by giving us a fairly objective way to partition data into inside and outside. Applying this method globally could help improve fundamental statistics about urbanization \citep{un2018, taubenbock2019new}.

It would also be interesting to use these methods, which give a robust way to define cities and regions, in constructing and assessing the various statistical laws which constitute the science of cities. We find that London has an outsized effect and sits at the centre of EW in terms of population and, especially, wealth, affirming its status as an outlier or even a `Dragon-King' \citep{pisarenko2012robust}. \cite{arcaute2015constructing} used a similar method of city boundary definition, but with arbitrary thresholds, and found that scaling laws were not robust to changes in that threshold. It would be very interesting to see if the thresholds determined here correspond to cities which fit on scaling curves. The relationship between the VP circle, scaling theory \citep{bettencourt2010unified} and other mathematical models of cities \citep{simini2012universal, pan2013urban} is also quite interesting, with the circles and profiles providing a new way to transform and analyse heterogeneous spatial data. In particular, the relationship of the ring model parameters $a_j$ to scaling exponents or other model parameters is unknown and understanding this connection is interesting future work.

\section*{Acknowledgements}
We thank Marcos Oliveira for helpful discussions and comments on the manuscript.

\bibliographystyle{elsarticle-harv} 
\bibliography{cas-refs}

\appendix
\section{Appendix: Robustness Checks}\label{sec:appgrid}

\begin{figure}[h]
    \centering
    \includegraphics[width=0.95\textwidth]{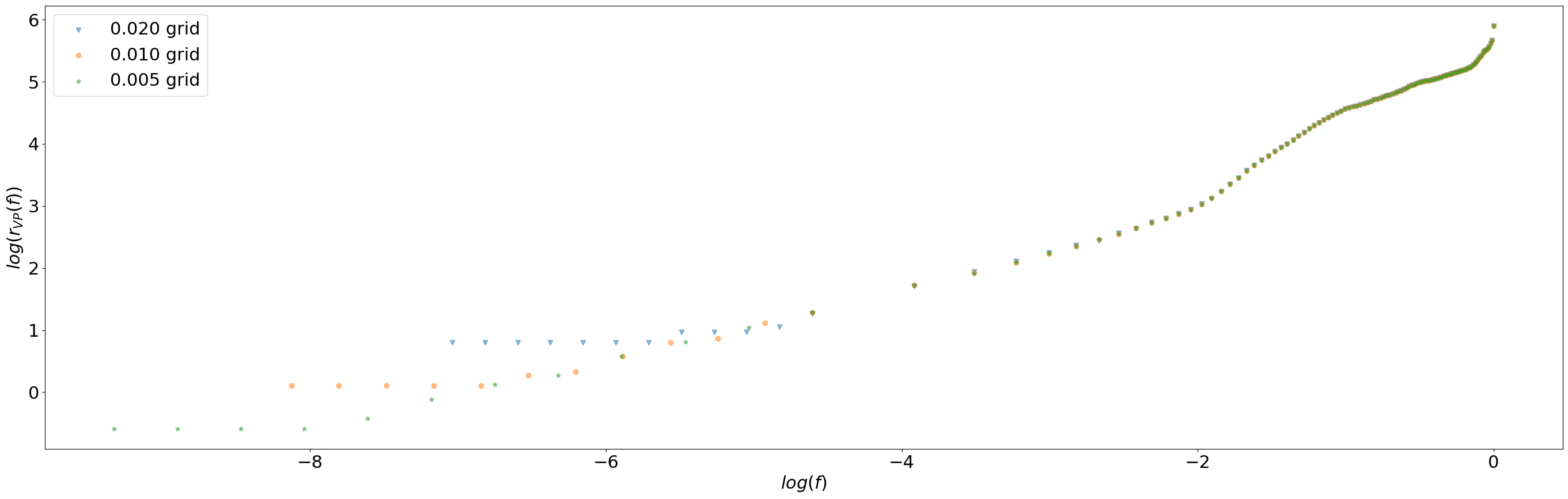}
    \caption{ $r(f)$ from population data using different grid resolutions. The grid boxes shows are approximately 2km, 1km and 0.5km on a side. After a value of about $\log f > -4$ the points completely overlap.}
    \label{fig:gridsize}
\end{figure}

\begin{figure}[h]
    \centering
    \includegraphics[width=0.95\textwidth]{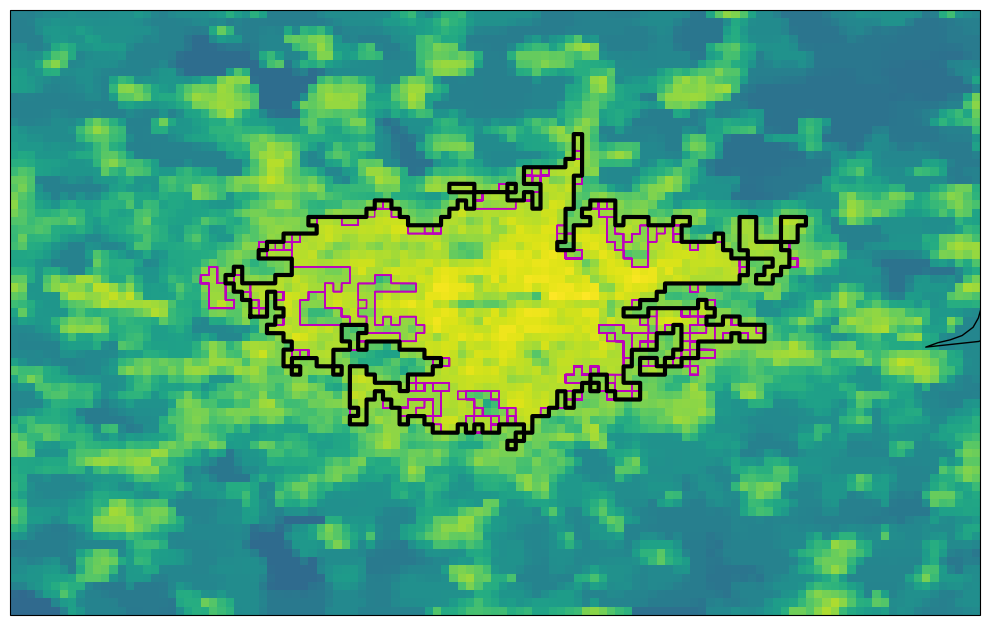}
    \caption{ London boundary derived from population data using different centres: $\pm 0.01$ degrees in latitude and longitude; different search radii: 2, 5, 8km; and different bounding box sides: 40, 50, 60km. The dark outline shows the parameters used in the main text, the pink lines are all of the boundaries found for any parameter combinations. }
    \label{fig:cityparams}
\end{figure}

Figure \ref{fig:gridsize} shows how the VP profile, the radius $r$ as a function of included population fraction, changes as the grid resolution is changed. Once the radius is large 
enough to include several grid boxes the profiles are identical. Figure \ref{fig:cityparams} shows how the city boundaries detected depend on the search parameters in the case of London 
population date. While there are some cases in the west of the city where quite low density areas are not included, generally the border is quite consistent. One would expect marginal 
areas on the edge of the city whose inclusion or exclusion is a close call with this, or any, method of boundary detection. One could develop this perturbation approach, for example, 
counting the fraction of the time a box is included inside or outside the city to try to determine a fuzzy boundary, but we leave this development for future work. Here we note that 
the boundaries found by our method are largely consistent and, where changes in the parameters change the boundaries, this is usually down to decisions on encompassing places like 
city parks e.g. Ealing Common, or other low density areas e.g. Northolt air base, within the city.

%\label{app1}


%%       \citet{<label>} ==> Lamport (1994)
%%       \citep{<label>} ==> (Lamport, 1994)
%%




%\begin{thebibliography}{00}


%\bibitem[Lamport(1994)]{lamport94}
%  \textit{\LaTeX: a document preparation system},
%  Addison Wesley, Massachusetts,

%\end{thebibliography}
\end{document}